\documentclass[twocolumn]{aastex62}

\usepackage{fontenc}
\usepackage{amssymb}
\usepackage{amsmath}
\usepackage{graphicx}
\usepackage{xspace}
\usepackage{natbib}

\usepackage{gensymb}
\usepackage{nccmath}
\usepackage{paralist}
\usepackage{txfonts}
\usepackage{pstricks}
\usepackage{todonotes}
\usepackage{url}
\usepackage{hyperref}
\hypersetup{
    linkcolor=black,
    filecolor=blue,      
    urlcolor=blue,
}

\begin{document}

\title{A Rapidly Varying Red Supergiant X-ray Binary in the Galactic Center}

%\email{agottlieb7@ufl.edu}
\author{Amy M. Gottlieb}
\affiliation{Department of Astronomy, University of Florida, Gainesville, FL, 32611, USA}

\author{Stephen S. Eikenberry}
\affiliation{Department of Astronomy, University of Florida, Gainesville, FL, 32611, USA}
\affiliation{Department of Physics, University of Florida, Gainesville, FL, 32611, USA}
\affiliation{University of Florida Research Foundation Professor, USA}

\author{Kendall Ackley}
\affiliation{Monash University, Scenic Blvd and Wellington Road, Clayton VIC 3800, Australia}
\affiliation{Department of Physics, University of Florida, Gainesville, FL, 32611, USA}

\author{Curtis DeWitt}
\affiliation{Department of Physics, University of California, Davis, CA 95616, USA}

\author{Amparo Marco}
\affiliation{DFISTS, EPS, Universidad de Alicante, Carretera San Vicente del Raspeig s/n, E-03690, San Vicente del Raspeig, Spain}
\affiliation{Department of Astronomy, University of Florida, Gainesville, FL, 32611, USA}

\begin{abstract}

We analyzed multiwavelength observations of the previously identified Galactic center X-ray binary CXO 174528.79–290942.8 (XID 6592) and determine that the near-infrared counterpart is a red supergiant based on its spectrum and luminosity. Scutum X-1 is the only previously known X-ray binary with a red supergiant donor star and closely resembles XID 6592 in terms of X-ray luminosity ($L_{\mathrm{X}}$), absolute magnitude, and IR variability ($L_{\mathrm{IR,var}}$), supporting the conclusion that XID 6592 contains a red supergiant donor star. The XID 6592 infrared counterpart shows variability of $\sim$0.5 mag in the Wide-field Infrared Survey Explorer-1 band (3.4 $\mu$m) on timescales of a few hours. Other infrared data sets also show large-amplitude variability from this source at earlier epochs but do not show significant variability in recent data. We do not expect red supergiants to vary by $\sim50\%$ in luminosity over these short timescales, indicating that the variability should be powered by the compact object. However, the X-ray luminosity of this system is typically $\sim1000\times$ less than the variable luminosity in the infrared and falls below the Chandra detection limit. While X-ray reprocessing can produce large-amplitude fast infrared variability, it typically requires $L_{\mathrm{X}}>>$\,\,$L_{\mathrm{IR,var}}$ to do so, indicating that another process must be at work. We suggest that this system may be a supergiant fast X-ray transient (SFXT), and that a large ($\sim$10$^{38}$\,ergs\,s$^{-1}$), fast ($10^{2-4}$\,s) X-ray flare could explain the rapid IR variability and lack of a long-lasting X-ray outburst detection. SFXTs are normally associated with blue supergiant companions, so if confirmed, XID 6592 would be the first red supergiant SFXT, as well as the second X-ray red supergiant binary.

\end{abstract}


\section{Introduction} \label{sec:intro}

The Galactic center (GC) is a region of extremely high stellar density. The central $2\degree\times0.8\degree$ of the Galaxy includes $\sim$1\,\% of stellar mass in the Galactic disk \citep{Launhardt2002}, and the central 50 pc contains $\sim0.1$\,\% of the total stellar mass and $\sim2$\,\% of the Galactic population of young, massive stars (e.g. \citet{Figer2004}). The GC environment also differs significantly from the rest of the Galaxy as the region is in the vicinity of Sgr A*, the 4.02 ($\pm0.16\pm0.04)\times10^6$ $M_\odot$ supermassive black hole (\citet{Boehle2016}, and references therein). The GC was originally seen in the X-rays as a very diffuse source, but high-resolution instruments (particularly Chandra), resolved most of this emission into over 10,000 X-ray point sources \citep{Muno2009}. Using X-ray hardness ratios, authors such as \citet{Muno2009} determined that the majority of the sources are in the GC (or beyond it), as opposed to in the foreground. While the number of X-ray sources here is very large, their nature remains largely unknown \citep{Wang2002,Muno2003}. This is partially because many of these sources are too faint to be seen at energies below 1.5 keV owing to the large amount of absorption between us and the GC. 

Chandra observations of the GC show populations of cataclysmic variables (CVs), late-type stars with active coronae, and unknown source classes likely to be compact object (white dwarf, neutron star, or black hole) accreting binaries \citep{Morihana2016}. NuSTAR (Nuclear Spectroscopic Telescope Array, \citet{NuSTAR2013}) observed the GC at higher energies (albeit lower angular resolution) and found many hard X-ray point sources \citep{Nustar2016} as well as an excess of unresolved hard (20-60\,keV) X-ray emission in the central few arcseconds \citep{Perez2015}, potentially produced by different populations such as millisecond pulsars (MSPs), quiescent low-mass X-ray binaries (qLMXBs), or intermediate polars (IPs) (which are magnetic CVs).

\citet{Hailey2018} report finding a `density cusp' of a dozen qLMXBs (which they suggest are black holes) within 1\,pc of the GC, compared to the nonthermal diffuse hard X-ray emission in the inner 8\,pc dominated by IPs \citep{Perez2015}. From the luminosity function and spatial distribution of the potential qLMXBs, \citet{Hailey2018} also inferred that there could be hundreds of such binary systems - and even more isolated black holes - in the GC. In another recent paper, \citet{Zhu2018} identified 1300 new Chandra sources, which they classify as both magnetic and nonmagnetic CVs, but claim that qLMXBs are only a minor population in the GC. However, all these interpretations rely almost solely on comparing the X-ray properties of these sources to previously studied major X-ray source populations in the broader Milky Way. Because the GC is such an unusual environment compared with the rest of the Milky Way, we expect to see new and interesting sources with atypical properties. In that case, improved understanding of the X-ray source population in the GC may depend on multiwavelength studies (particularly in the infrared, due to the extreme reddening toward the GC) to reveal the host star types, binary periods, mass functions, and other key properties of the systems. As we show here, XID 6592 appears to be one such highly unusual source.

\begin{figure}[ht]
  \includegraphics[trim={0 2.1cm 0 0},clip,width=\columnwidth]{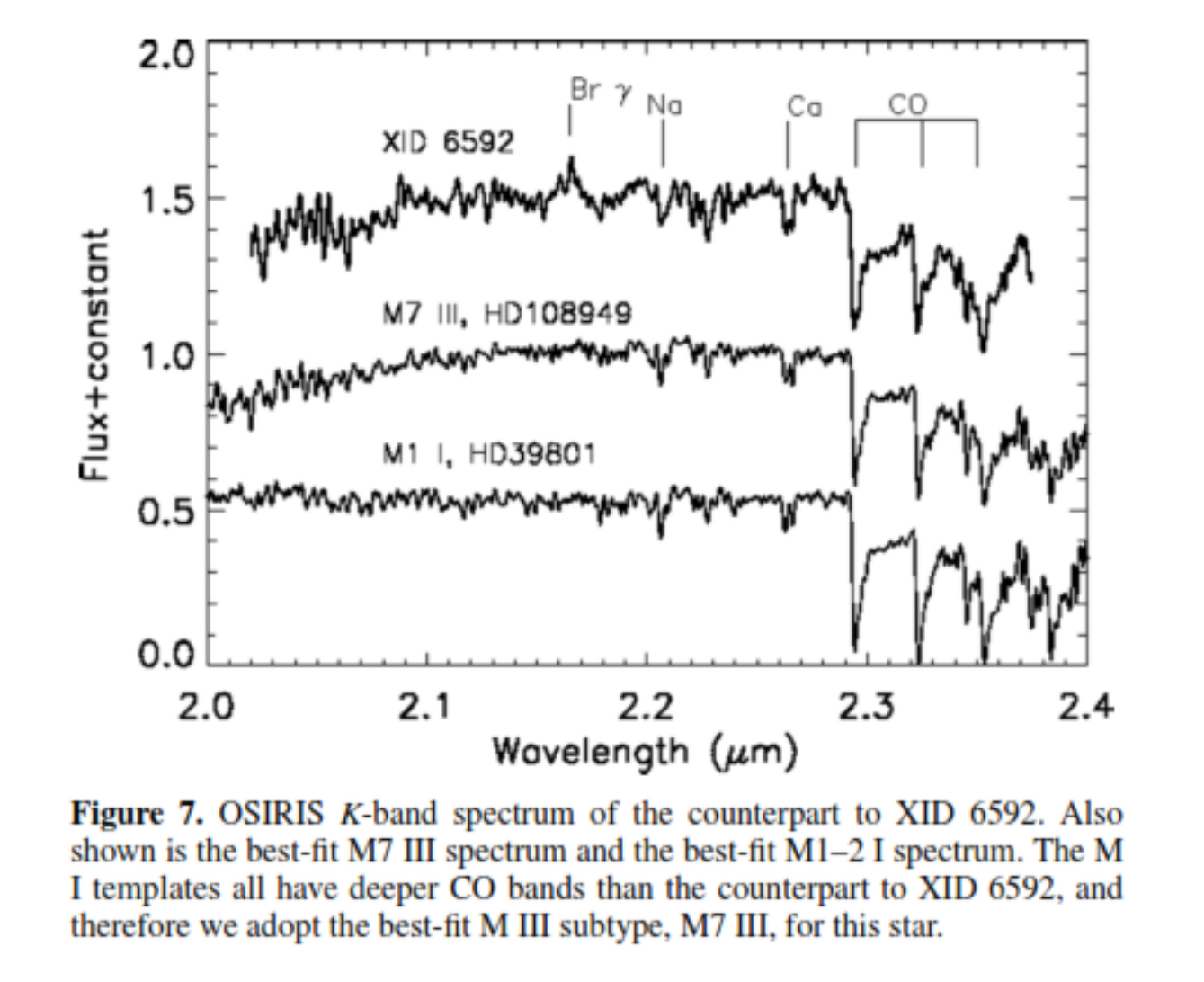}
  \caption{Figure 7 from \citet{DeWitt2013}. OSIRIS \textit{K}-band spectrum of the counterpart to XID 6592. Also shown is the best-fit M7III spectrum and the best-fit M1-2 I spectrum.   }\label{fig:xid_spec}
\end{figure}

\citet{DeWitt2013} identified possible near-infrared (NIR) counterparts to Chandra X-ray sources in the GC \citep{Muno2009} with high probabilities of being true counterparts as opposed to chance matches, where the two sources are at the same location on the sky but at different distances. They determined that XID 6592 has a true NIR counterpart and is therefore an X-ray binary. XID 6592 is located near the GC at R.A. = $17^{\mathrm{h}}45^{\mathrm{m}}28.79^{\mathrm{s}}$ and decl. = $-29^{\degree}09'42.8''$ ($l=359.\degree791396,\,b=-0.\degree091492$). A spectrum of the NIR counterpart taken with OSIRIS (Ohio State InfraRed Imager/Spectrometer; \citet{Depoy1993}) on the Cerro Tololo Inter-American Observatory 4 m telescope (shown in Figure \ref{fig:xid_spec}) shows a Brackett-$\gamma$ emission line that is characteristic of an accretion disk in an X-ray binary system, as normal stars do not typically show emission lines.  Other types of stars such as red supergiants (RSGs) and Mira variables have stellar winds that can produce this emission line, but these stars do not have significant X-ray luminosities, whereas XID 6592 does ($L_{\mathrm{X-ray}}\sim10^{33}$\,ergs\,s$^{-1}$; see Section \ref{subsec:chandra} and \ref{subsec:xmm}). The spectrum also shows broad CO absorption bands that are characteristic of late-type stars. \citet{DeWitt2013} determined this to be close to an M7-type star through spectral fitting. 

\begin{figure}[ht]
  \includegraphics[trim={4cm 1.2cm 3.5cm 0},clip,scale=0.5]{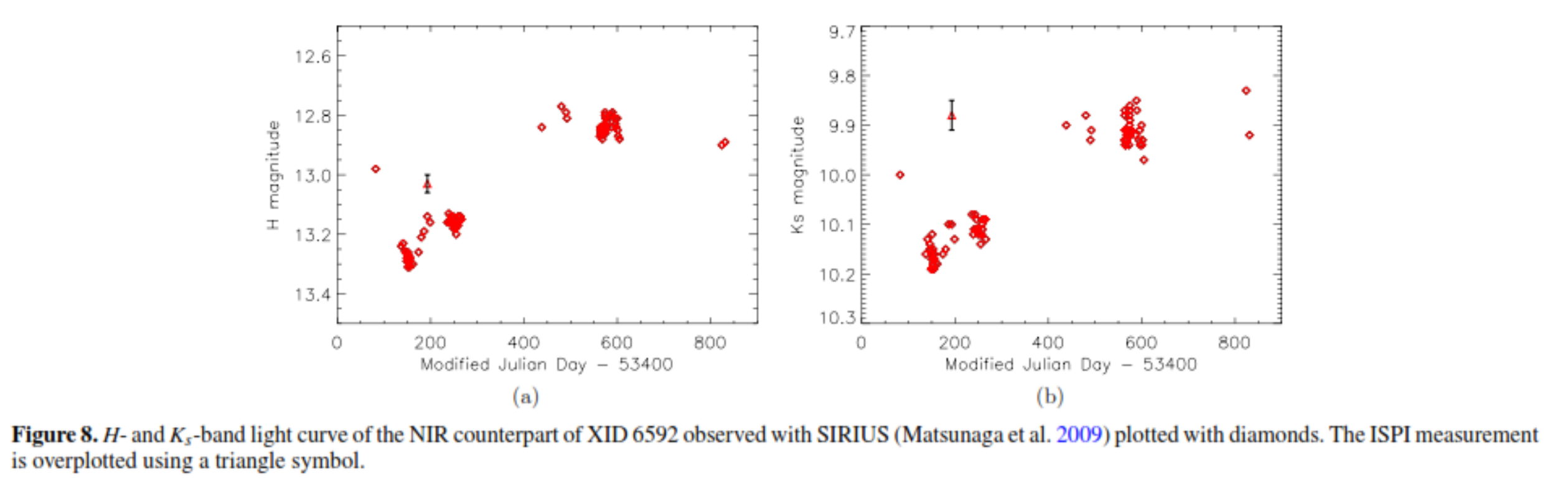}
  \caption{Figure 8 from \citet{DeWitt2013}. Data span 2005-2006. $H$- and $K_s$-band light curve of the NIR counterpart of XID 6592 observed with SIRIUS \citep{Matsunaga2009} plotted with diamonds. The ISPI measurement is overplotted using a triangle symbol.}\label{fig:isis_lc}
\end{figure}

\citet{Matsunaga2009} made photometric measurements of XID 6592 with SIRIUS (Simultaneous-color InfraRed Imager for Unbiased Surveys; \citet{Nagayama2003}) and ISPI (the Infrared SidePort Imager; \citet{vanDerBliek2004}) in 2005-2006 (Figure \ref{fig:isis_lc}), and concluded that this source is a long-period variable (LPV). However, they were unable to find the periodicity. The ISPI photometry from \citet{DeWitt2013} was an apparent outlier compared to the SIRIUS measurements, but they verified their analysis and interpreted this event as a possible flaring episode. This prompted us to search the archives for any other instances of NIR and X-ray variability in XID 6592.

In \S\ref{sec:obs} we describe the data and the data reduction procedure. In \S\ref{sec:results_disc}, we examine the light curves in both X-ray and infrared bands and discuss the results. Lastly, in \S\ref{sec:summary} we summarize the results of our temporal analyses.

\section{Observations and Data Analysis}\label{sec:obs}

We searched for and analyzed archival infrared images and X-ray data. Because this source is very bright in $K_s$-band, we searched not only $K_s$-band data, but also mid-IR data (such as Wide-field Infrared Survey Explorer [WISE] and Spitzer) for variability. We looked at both soft and hard X-ray missions, searching for any variability at softer energies and potential outbursts at harder energies.

\subsection{WISE Infrared Observations}\label{subsec:wise}

We searched the ALLWISE Multiepoch Photometry (MEP) Catalog \citep{Wright2010,2013allwise} in a 20$''$ region around the position of XID 6592 and found measurements for two epochs in 2010 March and 2010 September. To reduce the images and obtain magnitudes for this catalog, WISE observes a given location multiple times in all four filters at once to build up sensitivity (and also enabling a search for variability). The four bands (W1, W2, W3, and W4) are centered on 3.4, 4.6, 12, and 22 $\mu$m, respectively, with exposure times of 7.7 s for W1 and W2 and 8.8 s for W3 and W4. The field of view (FOV) is 47$\times$47 arcmin with a point-spread function (PSF) FWHM of 6$''$. The individual single exposures are combined in the Multiframe Pipeline to produce co-added images and a database of sources. The pipeline also performs a PSF chi-squared minimization in all bands simultaneously to obtain positions and mean fluxes for each detected source. The final fluxes are calculated by holding the position constant while fitting for the PSF amplitude.

\begin{figure}[ht]
  \centering
  \includegraphics[scale=0.68]{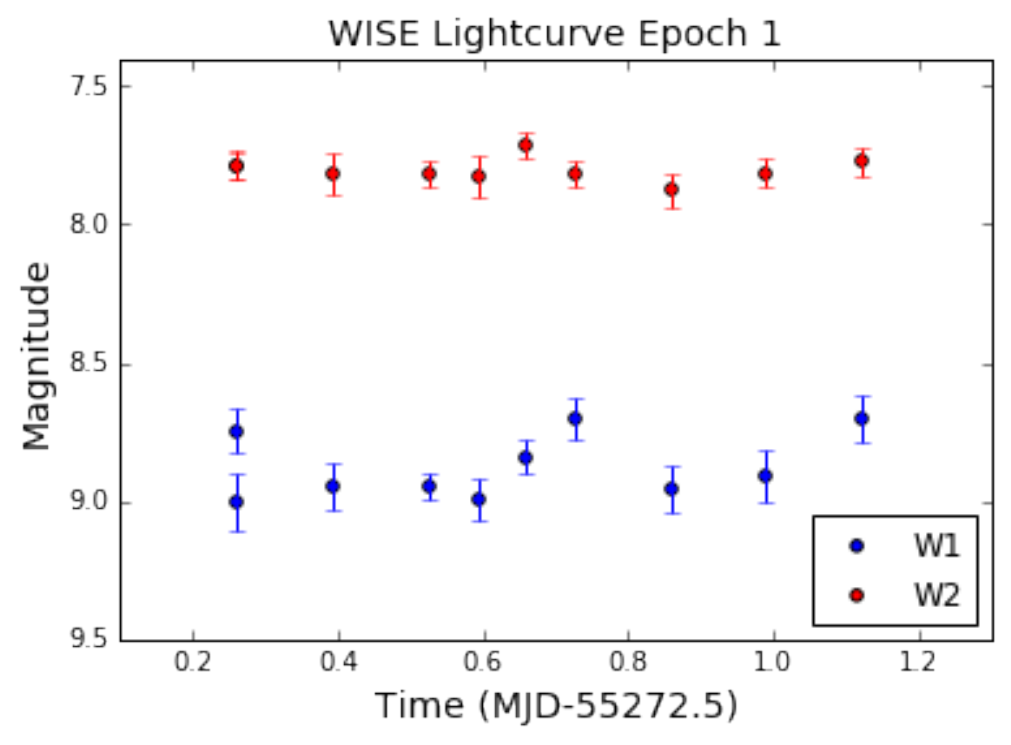}
  \caption{First epoch of the WISE 1 and 2 band light curves that occurred on 2010 March 17--18.}\label{fig:w1_lc1}
\end{figure}

\begin{figure}[ht]
  \centering
  \includegraphics[scale=0.68]{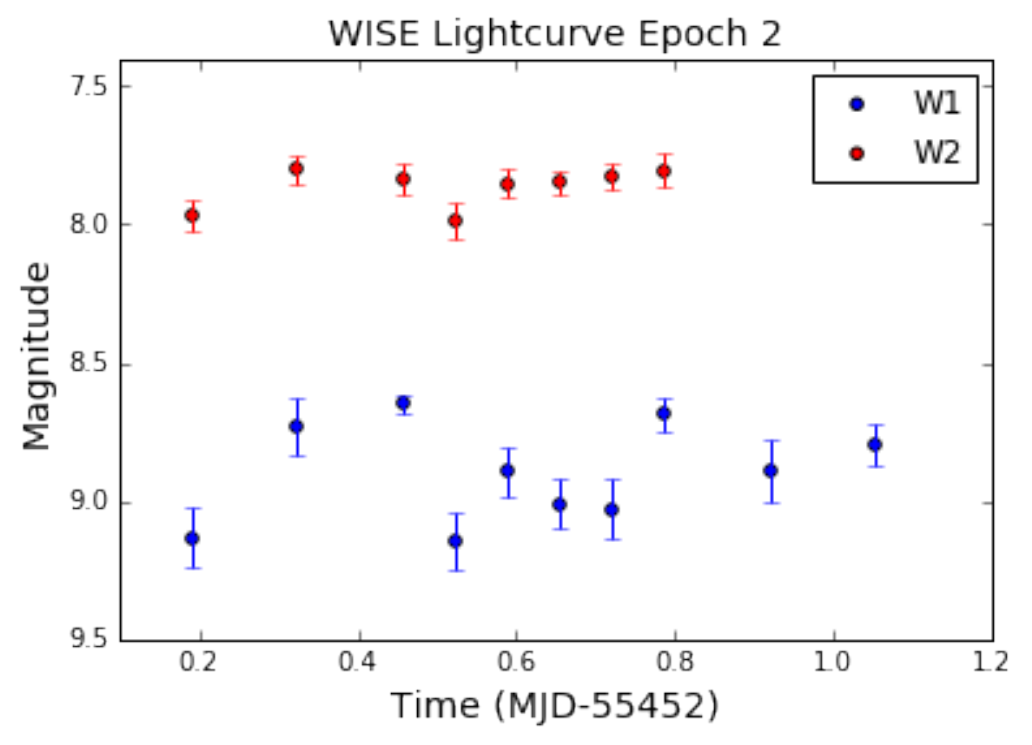}
  \caption{Second epoch of the WISE 1 and 2 band light curves that occurred on 2010 September 12--13.}\label{fig:w1_lc2}
\end{figure}

Using the data from the MEP Catalog, we show the final light curves for two epochs in the WISE 1 band (3.4\,$\mu$m) and WISE 2 band (4.6\,$\mu$m) in Figures \ref{fig:w1_lc1} and \ref{fig:w1_lc2}. We find significant variability particularly in the second epoch$-$ there is a change in magnitude of $\sim0.5$ mag in a few hours between 1.4 and 1.6 days, shown by the third and fourth points in Figure \ref{fig:w1_lc2}. There is no significant variability in the WISE 2 band light curves. The catalog contained data for the WISE 3 and 4 bands, but at these longer wavelengths blending becomes an issue, so these data are not very reliable. In the first two bands, blending is not an issue, indicating a higher likelihood of the reality of the observed variability.

To determine whether the variability we see is real and to quantify the variability of XID 6592, we calculated a reduced $\chi^2$ value and corresponding probabilities for three sets: the entire light curve as a whole and each epoch individually. We found that the W1 data were not consistent with a constant flux (reduced $\chi^{2} = 5.5$, equivalent to $6.5\sigma$). These results are discussed further in Section \ref{sec:results_disc}. 

\subsection{Spitzer Infrared Observations}

We searched the Spitzer Enhanced Imaging Products (SEIP) source list and the Infrared Science Archive (IRSA) for a 10 arcsec radius around the coordinates of XID 6592. We found both images and photometry of XID 6592 (SSTSL2 J174528.79-290942.9 in this catalog). The data were taken with the Infrared Array Camera (IRAC; \citet{Fazio2004}) on Spitzer \citep{Werner2004} in four different infrared bands (I1 = 3.6 $\mu$m, I2 = 4.5 $\mu$m, I3 = 5.8 $\mu$m and I4 = 8.0 $\mu$m) at resolutions of $\sim$2 arcseconds. We converted fluxes to magnitudes using the zero-points given in the \textit{Spizter} IRAC Instrument Handbook,\footnote{\url{https://irsa.ipac.caltech.edu/data/SPITZER/docs/irac/iracinstrumenthandbook/17/}}, and the results are shown in Table \ref{tab:spitzer}. 

The I1 measurement ($\sim$8.3\,mag) is brighter than the W1 measurement ($\sim$9\,mag); however, the Spitzer measurement may be averaged across multiple observations, and we do not know when any of the images were originally taken. This difference is not outside other IR variations seen (see below in Section \ref{subsec:vvv}). The I2 measurement is within the variations of W2, and we do not have another WISE measurement to compare with I3. In the I4 image, XID 6592 is visible but confused with the brighter background, which resulted in no photometric measurement in this band. 

\begin{table}[]
\centering
\caption{Spitzer IRAC Fluxes and Magnitudes}
\label{tab:spitzer}
\begin{tabular}{llll}
\toprule
 & I1 & I2 & I3 \\ \hline
Flux (Jy) & 0.1390(3) & 0.1426(3) & 0.1524(4) \\
Zero point (Jy) & 280.9$\pm$4.1 & 179.7$\pm$2.6 & 115.0$\pm$1.7 \\
Magnitude & 8.264$\pm$0.015 & 7.751$\pm$0.016 & 7.194$\pm$0.016 \\ \hline
\end{tabular}
\end{table}

We also searched for XID 6592 in MIPSGAL data. MIPSGAL is a Galactic plane survey using the Multiband Infrared Photometer \citep{Rieke2004} on Spitzer. The data were taken at 24$\mu$m with a resolution of 6$''$. There was no photometry available for XID 6592 in this band, as the image at the location of XID 6592 shows an excess nearby, but XID 6592 is not visible. 

\subsection{VVV NIR Observations}\label{subsec:vvv}

\begin{figure}[ht]
  \centering
  \includegraphics[scale=0.7]{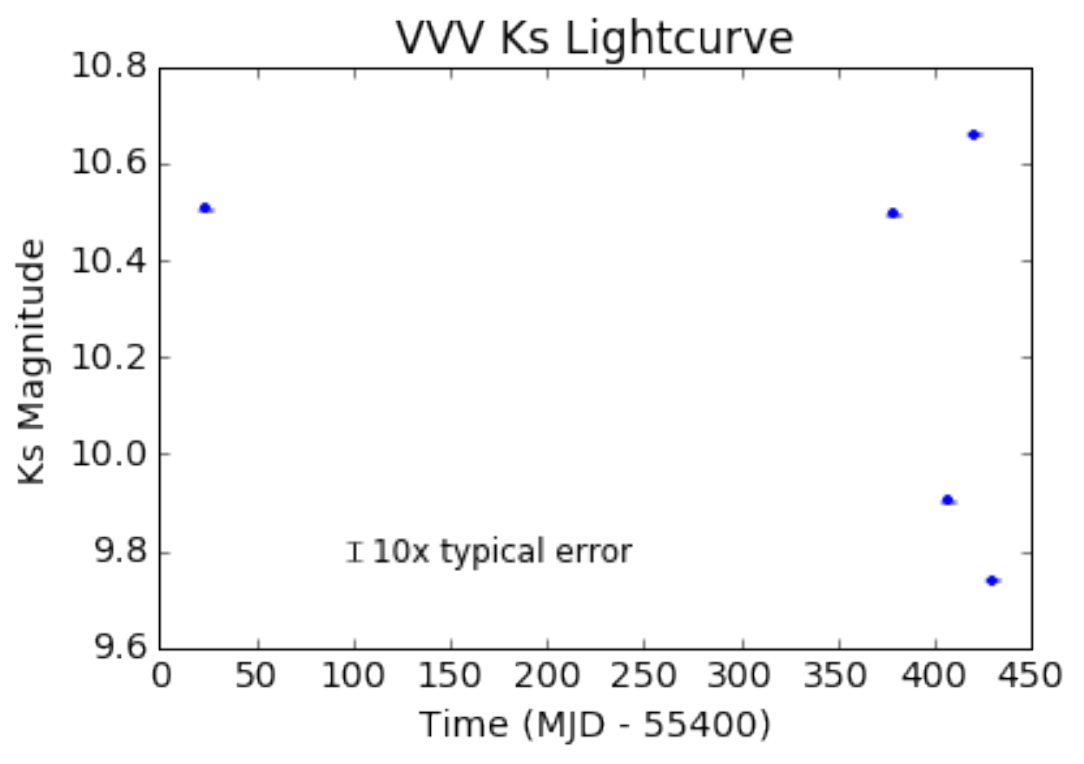}
  \caption{Full VVV light curve from catalog (2010/8 - 2011/10). Errors are all approximately 0.002 mag.}\label{fig:vvv_lc}
\end{figure}

The VVV (VISTA Variables in the Via Lactea; \citet{Minniti2010}, \citet{Saito2012}) data were taken on the 4 m VISTA telescope at Cerro Paranal Observatory in Chile in the $JHK_{s}YZ$ bands with VIRCAM, the VISTA InfraRed CAMera. VIRCAM has 16 chips that cover a 1.65$\degree$ diameter FOV \citep{Dalton2006,Emerson2006}. VVV used exposure times of 16 s on the GC, where the typical FWHM is $\sim$0.5$''$ and saturation occurs at $\sim11$ mag. By searching for archival VVV data of XID 6592, we found both cataloged magnitudes (by searching by IAU name, J174528.78-290942.77) and raw pawprint images containing XID 6592 (by searching for 20$''$ radius around the coordinates of XID 6592 and by region, b333). There are only a few data points available in the ESO VVV Multi-Epoch $K_s$ Band Photometry in the Via Lactea catalog (Source ID = 515535440768) that cover 2010--2011, shown in Figure \ref{fig:vvv_lc}. In the VVV catalog data, we see a change of $\sim$1 mag in the Ks band over approximately 10 days (as compared to the few-hour timescale in the WISE 1 data at 3.4 $\mu$m, at a different epoch).

We found over 500 raw images in the VVV archives containing XID 6592 that cover 2010--2017. However, XID 6592 saturates in many of the images, so we developed a method for regaining the information lost as a result of saturation. We describe the method here briefly (see Gottlieb et al. in preparation, for a detailed description.). In short, we use a set of Zernike polynomial base functions to model the unsaturated stellar PSFs in the image and create a normalized model PSF. We then fit the model PSFs to the data on XID 6592 after applying a mask (to ignore the bad, saturated pixels) to find the best-fit PSF amplitude and thus obtain the flux of the saturated star.

To create the model PSFs, we first run Source Extractor on the detector chip that contains XID 6592 to obtain a list of candidate sources, and then we filter out bad sources. The code examines all the stars in the catalog and selects stars that:

\begin{enumerate}
\item are not too close to the edge of the image (a 27$\times$27 pixel stamp centered 
on the star must be completely within the image); 
\item are point-like as opposed to elliptical (Source Extractor uses the pixel scale and seeing FWHM as inputs to a Neural Network that is trained to discriminate between stars and galaxies and outputs number between 0 (very elliptical) and 1 (perfectly circular); our stars must have a minimum value of 0.9 for this shape requirement);
\item have fluxes greater than the minimum flux requirement of 50,000 ADU; and
\item have zero flags from Source Extractor (these flags show which objects are bad in some way, e.g., blended, saturated, close to the edge, or close to another bright star).
\end{enumerate}

We ultimately want nine model PSFs with different subpixel centers in a $3\times3$ grid to obtain fractional pixel accuracy consistent with Source Extractor. Once we have the final filtered catalog of nonsaturated stars, they are assigned a position in the 3$\times$3 PSF tile based on the subpixel center of the star. Finally, we apply three iterations of $3\sigma$ clipping to the set of stars in each tile, take the average of stars that are left, and normalize each tile (divide by the sum of the average) to get the final model PSF for each tile with different sub-pixel centers. 

In order to determine how well the model PSF fits the data (e.g. XID 6592 and other stars in the image), we first applied the fitting to both saturated and nonsaturated stars without any mask and compared it to aperture photometry. We selected over 200 relatively isolated stars (noting that the GC is a very crowded region) with a large range in brightness with which to test this method. We initially filtered these stars by location (they should not be too close to the edge of the image as described above), and then found their centers and obtained stamps and subpixel centers. These stars are later filtered by flux to select those that lie in a reliably linear regime. For each star, we fit each of the nine model PSFs by first unraveling the star and the model PSF, and then finding the best scale and offset to minimize the $\chi^2$ (equation given below), defined as the difference between the data and the model that includes a scale factor and offset: 

\begin{ceqn}
\begin{align}
\chi^2 = \sum (D - (s \times M + o) )^2
\end{align}
\end{ceqn}

where $D$ is the unraveled star data, $s$ is the scale factor, $M$ is the unraveled model PSF, and $o$ is the offset.  We then take the PSF that resulted in the lowest $\chi^2$ (closest to zero). 

To obtain the final flux from the PSF fitting, the best-fitting scale and offset are applied to the best-fitting PSF, and we perform aperture photometry on this scaled-up model (i.e. applying the scale and offset to the model: $s\times M + o$) with the Python aperture photometry package, with the background subtracted as follows:

\begin{equation}
\centering
F = A - B \times N 
\end{equation}

where $F$ is the final flux, $A$ is the aperture photometry calculated from the Python package, $B$ is the median background within the annulus, and $N$ is the number of pixels within the aperture, which is equal to $\pi\,\times\,r$.

For each image, we determine the appropriate mask for XID 6592 and ignore all of these ‘bad’ pixels when fitting by setting them to NANs (Not A Number). These bad pixels are produced by ``supersaturation", where the source reaches near saturation in the first read of the IR detectors. In correlated double sampling and similar readout schemes, this results in a near-zero or even negative apparent flux.

Once we determine the appropriate mask for XID 6592, we apply it to all the stars that passed the filtering by flux above and repeat the fitting with the masked pixels set to NAN. From the fit process, we obtain the best-fitting PSF model (from the 9 in the $3\times3$ grid) and corresponding scale and offset. We then perform aperture photometry on the best-fitting unmasked, scaled-up model to get the final masked PSF-fitting photometry. 

Finally, to calculate the final aperture photometry flux of XID 6592, we plug in the masked PSF-fitting photometry value into the masked linear fit. To calculate the error bars on the magnitude of XID 6592, we take the RMS of the percent differences of the 20 brightest stars. We only use the 20 brightest stars because as flux increases, the scatter should decrease, 
so the scatter of the brighter stars is a better representation of the error in XID 6592, which is also bright.

After obtaining fluxes for XID 6592 in each of the images, we obtain a raw unbinned light curve. We need to account for other effects that could cause variability, such as atmospheric transmission. We correct for this by dividing the flux of XID 6592 by the sum of fluxes of other fainter, relatively isolated, unsaturated stars in the same image. Because the unsaturated stars are fainter, we select many of them to boost the signal-to-noise Ratio (S/N). The same set of stars must be used for all images for consistency. This was an issue for these particular observations because there is no set of stars that appears in every image owing to the VVV dithering pattern. In some images, XID 6592 is on the far right of the chip, and in others on the far left. Therefore, we split the image into quadrants and selected 20 stars in each quadrant.

\begin{figure}[ht]
  \includegraphics[width=\columnwidth]{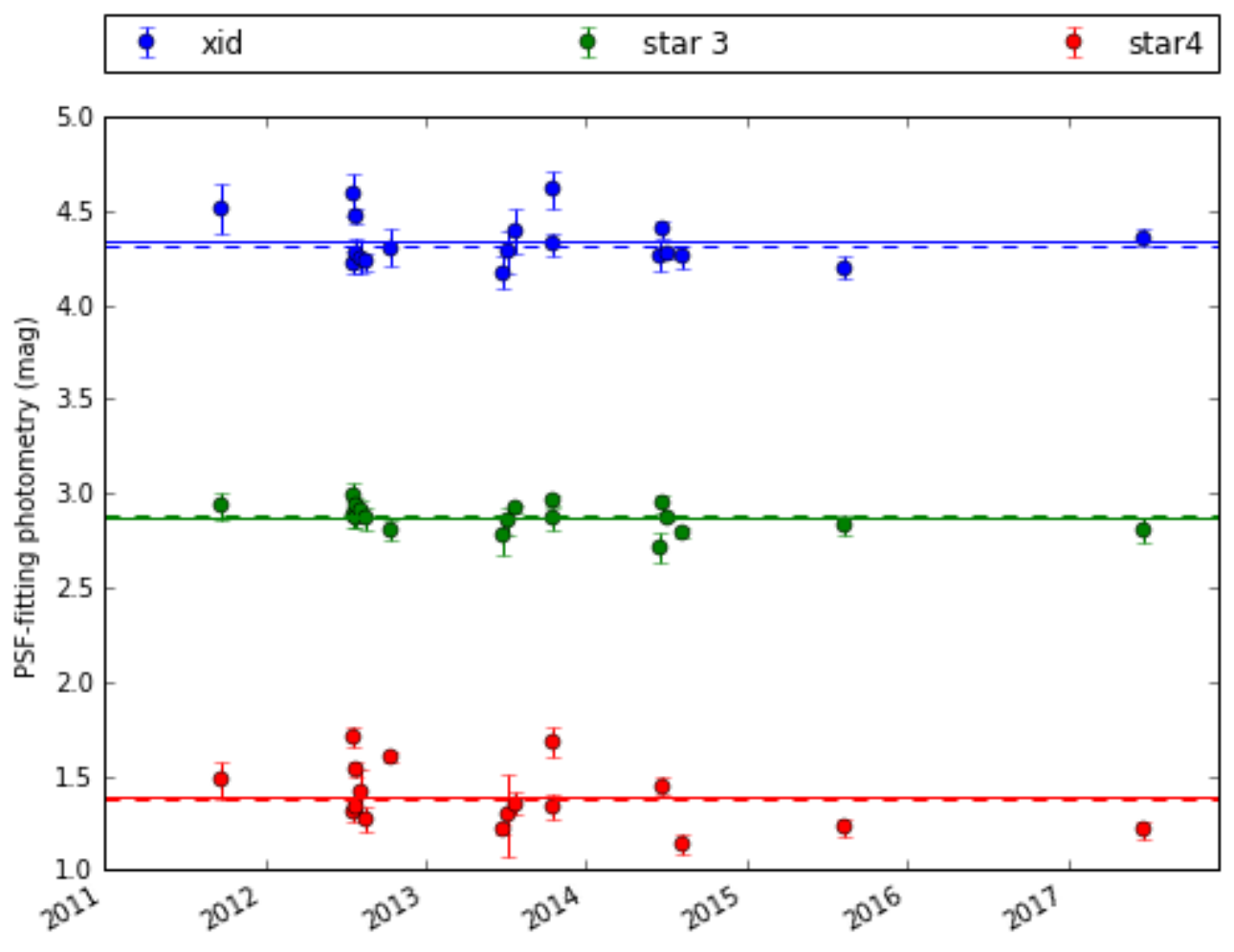}
  \caption{light curve of XID 6592 in blue (top) and two reference stars in green (middle) and red (bottom) obtained by using the PSF-fitting photometry method (aperture photometry on the fitted PSF model). The reference stars were shifted for clarity. The dashed and solid lines are the averages with errors taken into account and without errors taken into account, respectively.}\label{fig:vvv_lc2}
\end{figure}

Once we corrected for atmospheric effects, we then averaged all of the data points in one day together to produce the final light curve of XID 6592 shown in Figure \ref{fig:vvv_lc2} as the uppermost light curve. We performed the same analysis on two other field stars similar in brightness to XID 6592 (described below), shown in the middle and bottom light curves. To calculate the errors on the binned data, we use the larger of (1) the scatter of the data points within each bin (all on the same day) or (2) the average of the error bars of all the data points within each bin. 

We then picked two relatively isolated field stars (denoted star 3 and star 4) of similar brightness and repeated this entire process. Star 3 has $K_s$ = 10.41 $\pm$ 0.03 mag and star 4 has $K_s$ = 9.83 $\pm$ 0.03 mag (from the Two Micron All Sky Survey [2MASS]; \citet{Skrutskie2006}), compared to XID 6592, which had $K_s$ = $9.893\pm0.039$ mag in 2MASS. The final light curves of all three stars are shown in Figure \ref{fig:vvv_lc2}. To determine whether XID 6592 is more variable than these other two reference stars, we calculated the reduced $\chi^{2}$ for each star ($\sim$3 for XID 6592, $\sim$3 for star 3, and $\sim$13 for star 4). The errors in our measurements are likely dominated by systematics; however, it would not change our conclusion that XID 6592 is not significantly variable compared to the reference stars. The $\chi^{2}$ for star 4 is dominated by a few points that are far from the mean and also have small error bars. While the exact cause of this behavior is unclear, the conclusion remains the same: we do not see any significant variability in XID 6592 in this data set, which has many defects and problems.

\subsection{CIRCE NIR Observations}\label{subsec:circe}

On 2016 July 22 and 23, we observed XID 6592 with the NIR instrument CIRCE (Canarias InfraRed Camera Experiment; \citet{Eiken2018}) at the Gran Telescopio Canarias (GTC) 10.4 m telescope. The exposures were dithered and had 3\,s exposures at each position in the $K_s$ band. We reduced the images with SuperFATBOY, which performs normal reduction methods including dark and sky subtraction, flat-fielding, removing cosmic rays, and aligning and stacking the images, as well as some other finer corrections such as masking bad pixels, applying a linearity correction, and deboning the image to remove the underlying herringbone structure \citep{Warner2012,Warner2013}.

\begin{figure}[ht]
\centering
  \includegraphics[scale=0.28]{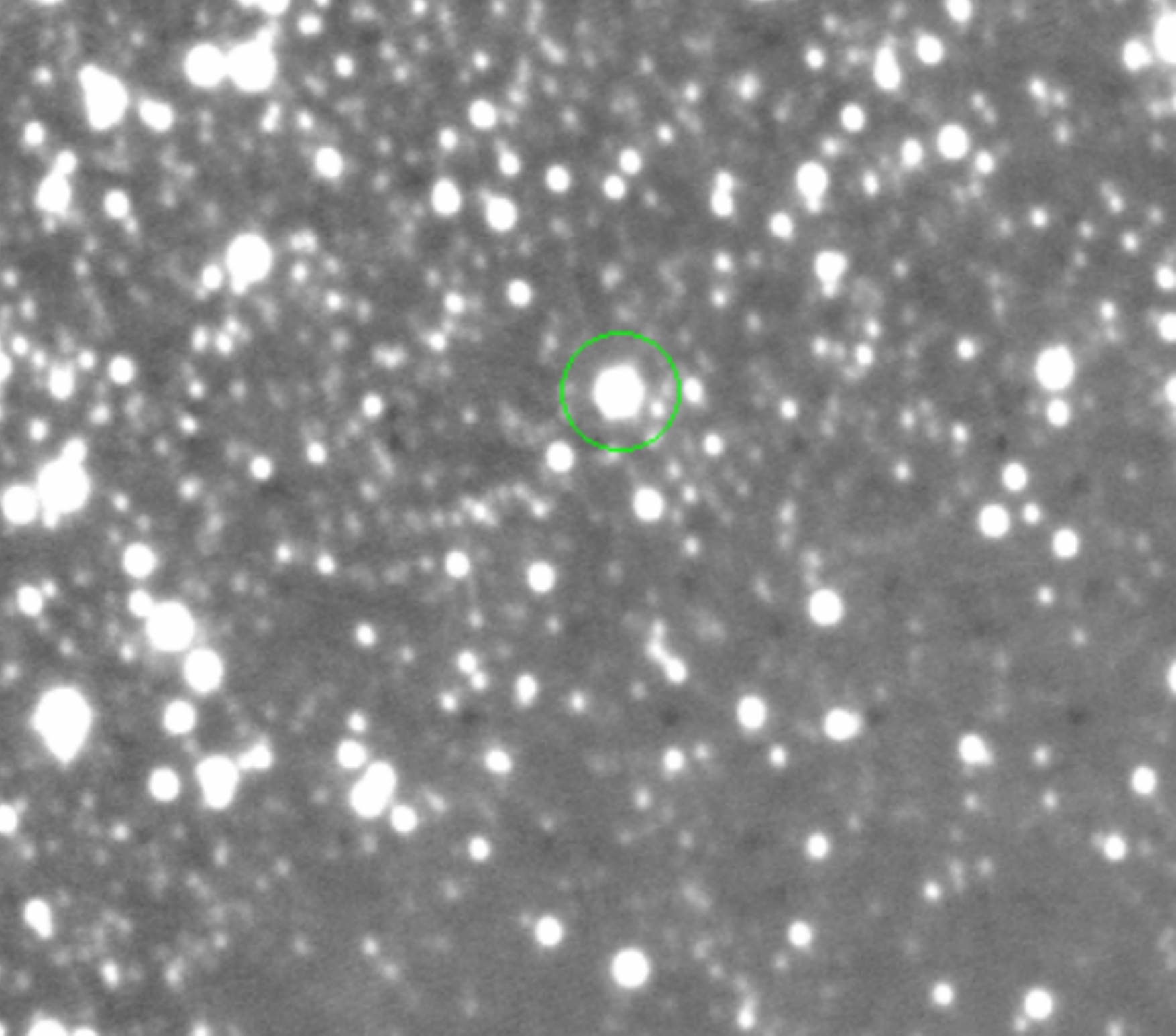}
  \caption{Final reduced CIRCE image of the NIR counterpart to XID 6592, shown in the green circle (radius = 3$''$).}\label{fig:circe}
\end{figure}

We show an example of one of the final processed images in Figure \ref{fig:circe}. We performed aperture photometry on XID 6592 and 10 other fainter stars in the image using the Python package \textbf{aperture photometry} within \textbf{photutils}. We compared these with the 2MASS catalog magnitudes by fitting a line to aperture photometry versus 2MASS photometry to determine the instrumental magnitude offset. After fitting the line, we calculated the percent difference between the calculated magnitudes and the fitted line and then took the RMS of this value as the error bar for XID 6592, as we did for the VVV analysis. We calculated the magnitude of XID 6592 to be $9.705\pm0.017$ mag and $9.692\pm0.016$ mag on 2016 July 22 and 23, respectively.

\subsection{Chandra X-Ray Observations}\label{subsec:chandra}

\begin{figure}[ht]
\centering
  \includegraphics[scale=0.46]{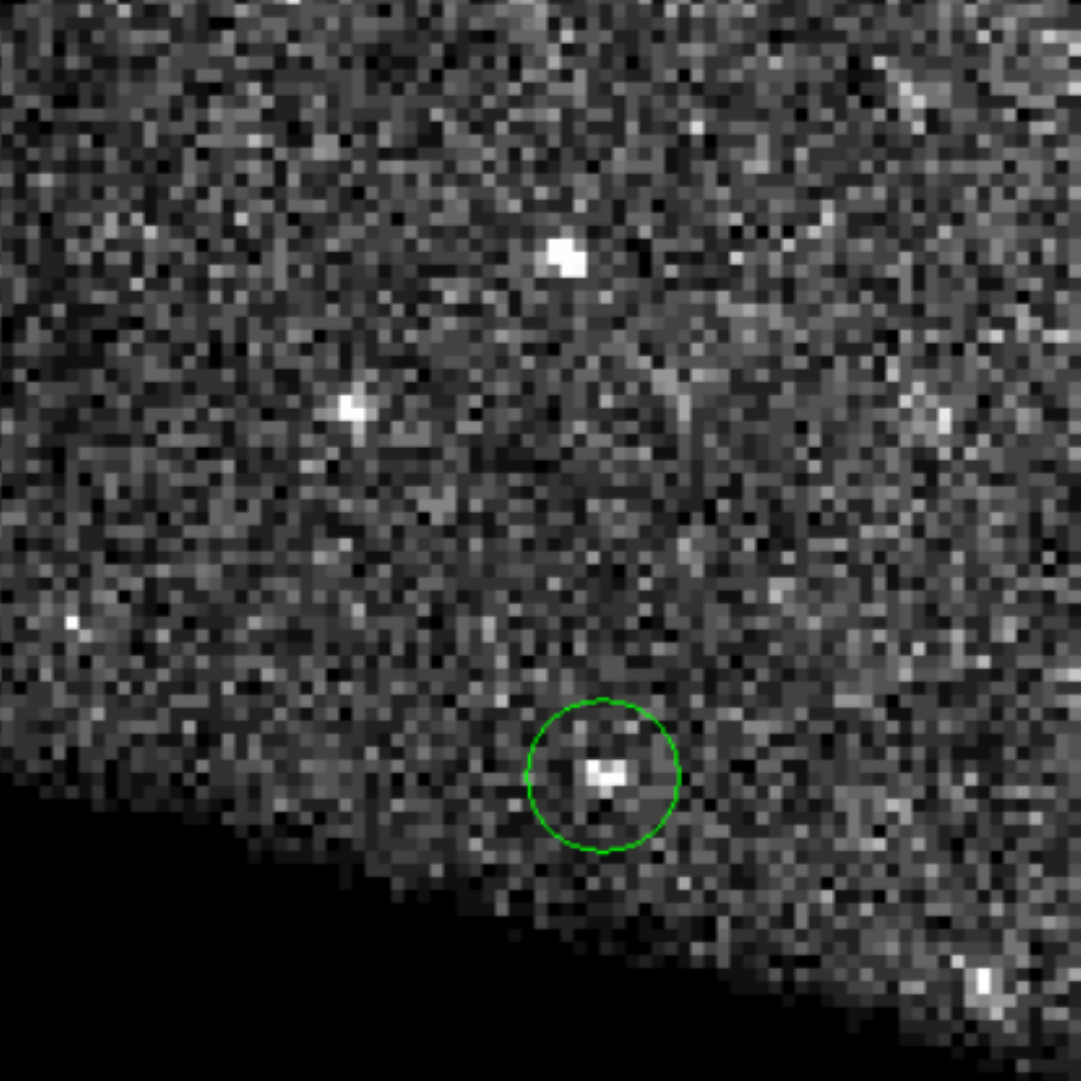}
  \caption{Chandra image where XID 6592 is within the green circle, which is 20$''$ in radius. }\label{fig:chandra_img}
\end{figure}

By searching the Chandra Data Archive for observations covering the position of XID 6592, we compiled a list of X-ray observations of the GC taken by the Advanced CCD Imaging Spectrometer (ACIS) on Chandra \citep{Weisskopf1999, Weisskopf2002}. ACIS has a very high angular/spatial resolution of 0.5--1 arcsecond compared to other X-ray detectors and a relatively small (0.5$\degree$ in diameter) FOV. It is an imager that detects individual photons and records their positions, energies, and arrival times. We obtained high-level products in the form of FITS files processed by the standard pipeline from the archive, which only contain position and counts information, not energy and arrival time.

We obtained images containing the R.A. and decl. of XID 6592 in the field of view using the FOVFiles tool. Chandra detected XID 6592 in the following observations: ObsID 1561, 2291, 2293, 4683, 4684, 5950, 5951, and 7037, with nondetections in 5950, 5953, 7037, 7557, and 9173 (but its position was located on a chip). An example of a Chandra ACIS image containing XID 6592 is shown in Figure \ref{fig:chandra_img}. 

We used CIAO tools and the following process to analyze the images and obtain counts s$^{-1}$. First, we used \textbf{celldetect} to find sources in the image and obtain the dimensions of the source and background regions. While \textbf{wavdetect} is more commonly used in crowded regions like the GC, XID 6592 is in a relatively isolated area near the GC. Therefore, the use of \textbf{celldetect} is acceptable. \textbf{Celldetect} calculates the S/N of `source' counts to background counts at each place where a sliding square cell the size of the instrument PSF is placed. The source is recorded as a candidate if the S/N is above a detection threshold. Once the source was detected and we obtained the dimensions of the source and background regions, we used the \textit{analysis} tool to obtain the number of counts within each region. There were a total of 1693 counts across 13 observations, with an average of 0.003 counts\,s$^{-1}$. Then, we used the \textbf{aprates} tool to calculate the rate of the source  and the bounds on the rate given the number of source and background counts, the area of the regions containing the number of source and background counts, and the exposure time of the image. Finally, we converted the rates to unabsorbed 0.2--8\,keV flux using the PIMMS simulation tool with N$_{\mathrm{H}} = 9.6\times10^{22}$\,cm$^{-2}$ (which corresponds to an extinction, $A_{\mathrm{V}}$, of $\sim53$\,mag) and a power law with $\Gamma$=$1.5$ as used in \citet{DeWitt2013}. Given the large bandpass, using $\Gamma$=0.5 and 2.5 only changes the flux by a factor of 3, and using an $N_{\mathrm{H}}$ corresponding to $\sim40$\,mag of extinction in the $V$ band only results in 10$\%$ lower fluxes. 

\begin{figure}[ht]
  \includegraphics[trim={5.9cm 12cm 6cm 11.8cm}, clip,scale=0.9]{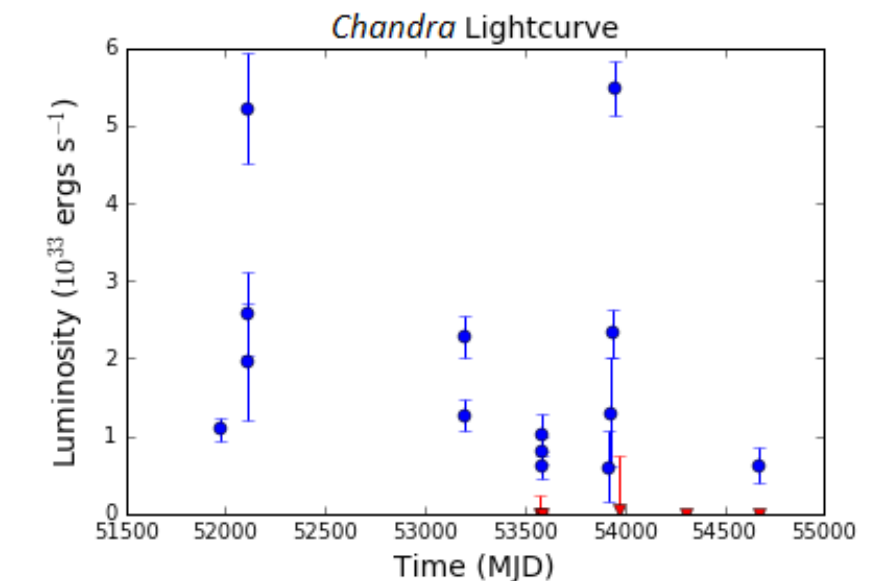}
  \caption{Full Chandra light curve (2001 April - 2008 March) assuming a power law with $\Gamma$=$1.5$ and $N_{H} = 9.6\times10^{22}$\,cm$^{-2}$. The luminosity was calculated using a GC distance of $8.12\pm0.03$\,kpc. Red points are upper limits where the source was not significantly detected.}\label{fig:chandra_lc}
\end{figure}

\begin{figure}[ht]
  \centering
  \includegraphics[scale=0.7]{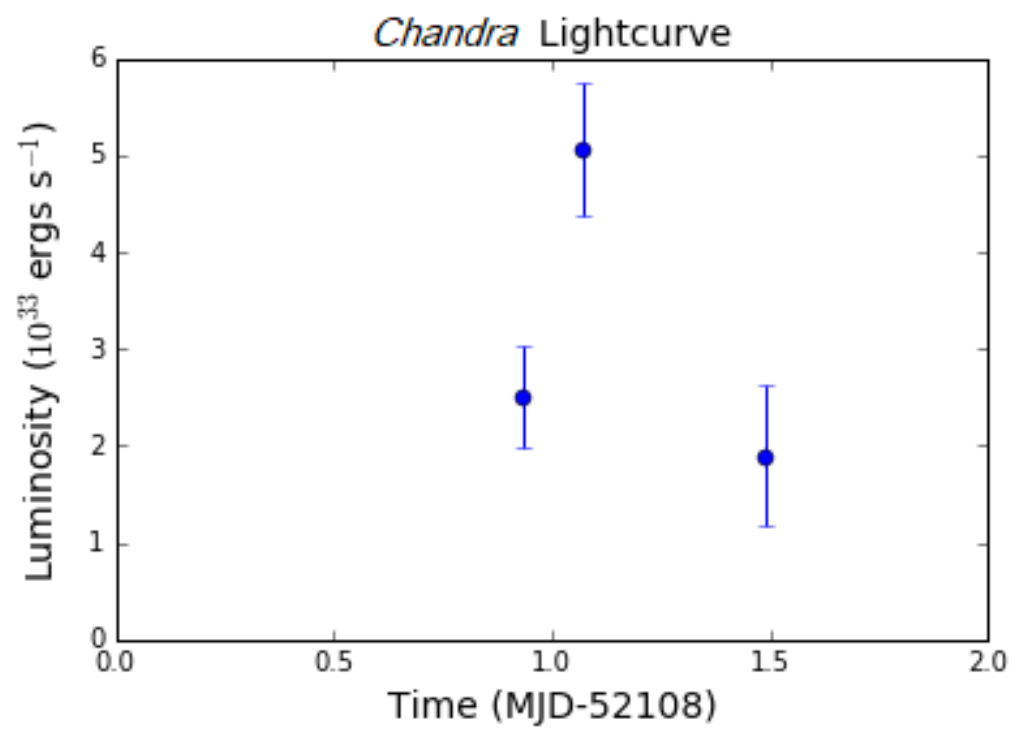}
  \caption{Zoom-in on the short brightening event in Chandra light curve.}\label{fig:chandra_lc1}
\end{figure}

\begin{figure}[ht]
  \centering
  \includegraphics[scale=0.7]{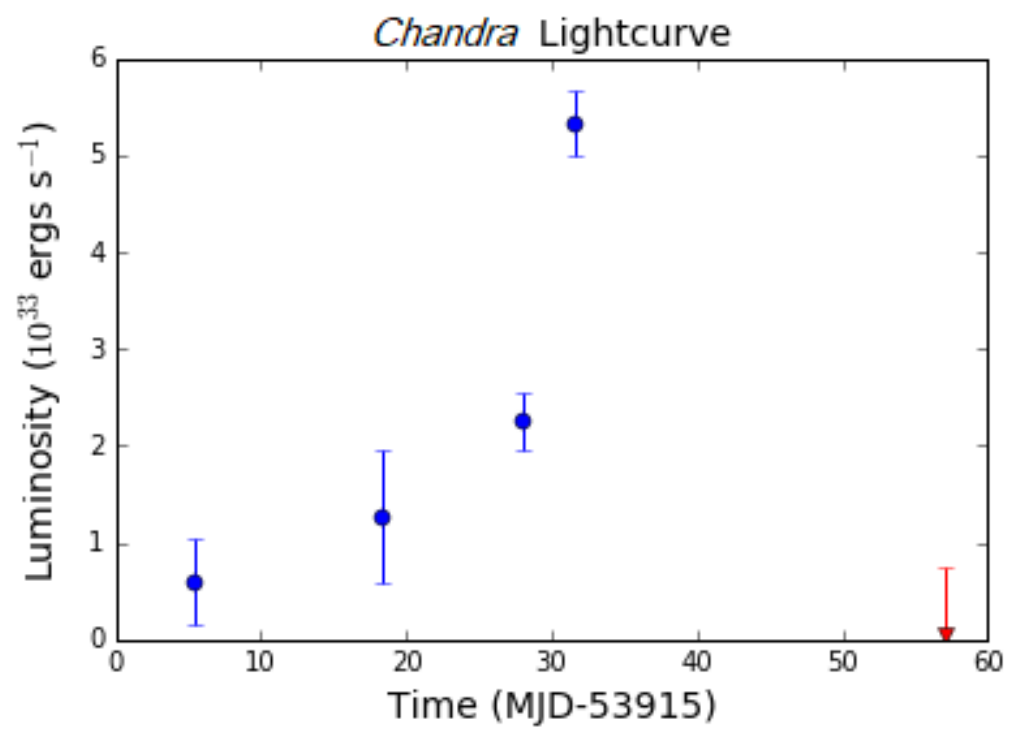}
  \caption{Zoom-in on the long brightening event in Chandra light curve.}\label{fig:chandra_lc2}
\end{figure}

We converted flux to luminosity using a GC distance of 8.12$\pm$0.03\,kpc \citep{Gravity2018} and then used these data to create light curves. We assume that XID 6592 is at the GC because its location and extinction are consistent with the GC. There is also a much higher density of X-ray binaries in the GC than outside of it, making the probability of XID 6592 being in foreground very low. The full light curve is shown in Figure \ref{fig:chandra_lc}, and zoomed-in views of the brightening events are shown in Figures \ref{fig:chandra_lc1} and \ref{fig:chandra_lc2}. It is possible to obtain a spectrum of sources in the center of the image on the chips where events (containing position, arrival time and energy as opposed to just position and number of photons) are recorded. However, because XID 6592 is relatively faint in the X-rays, the spectrum would have a low S/N. Also, we were only interested in the brightness, so we did not analyze any event files. 

\subsection{XMM X-Ray Observations}\label{subsec:xmm}

\begin{figure}
  \centering
  \includegraphics[trim={3.7cm 10cm 3.7cm 10cm}, clip,scale=0.55]{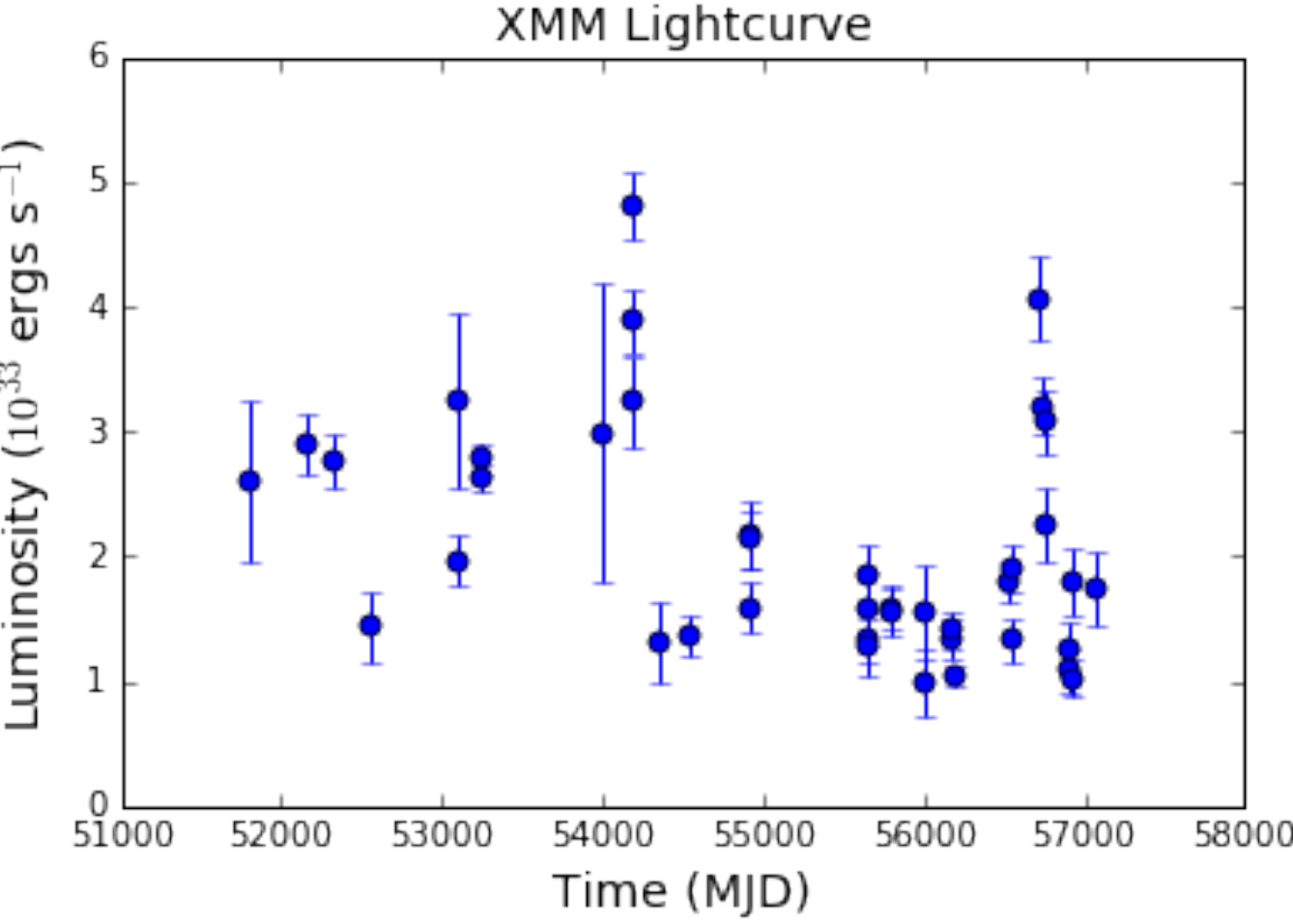}
  \caption{Full XMM light curve of XID 6592 (2000--2015) in the 0.2--12\,keV band using a distance of $8.12\pm0.03$\,kpc.}\label{fig:xmm_lc}
\end{figure}

We obtained fluxes of XID 6592 from 2000 to 2015 in the 0.2--12\,keV band by searching the 3XMM-DR8 Catalog \citep{Rosen2016} within the XMM-Newton (X-Ray Multi Mirror Mission; \citet{Turner2001, Struder2001}) Science Archive by R.A./decl. The data were taken with the European Photon Imaging Camera (EPIC), which is made up of three CCD cameras, one PN (back illuminated) camera and two Metal Oxide Semi-conductor (MOS) cameras, and has a FOV of 30$'$. It also has moderate angular resolution, with a 6$''$ FWHM PSF. This is a much larger FOV but a much worse angular resolution compared with Chandra. The raw data were run through the Pipeline Processing System (PPS) and final fluxes for each observation were made available in the catalog. We used a GC distance of $8.12\pm0.03$\,kpc to convert to luminosity. The final light curve is shown in Figure \ref{fig:xmm_lc}. 

\subsection{Other X-Ray Missions}\label{subsec:xray_obs}

We searched for other X-ray observations and all sky surveys containing XID 6592, including the Rossi X-ray Timing Explorer \citep[RXTE;][]{RXTE}, which covers the energy range 2--12\,keV, the INTErnational Gamma-Ray Astrophysics Laboratory \citep[INTEGRAL;][]{Winkler2003}, which covers the energy range 3\,keV--1\,MeV, the Monitor of All Sky X-ray Image \citep[MAXI;][]{Matsuoka2009}, which covers the energy range 2--20\,keV, and the Burst Alert Telescope (BAT) and the X-ray Telescope (XRT) on board the Neil Gehrels Swift observatory \citep{Gehrels2004,Barthelmy2005}, which cover the energy ranges 15--150\,keV and 0.3--10\,keV, respectively. We searched across all time and in particular around the time of the WISE 1 fast variability (2010 September 12–13), but we did not find XID 6592 in any of the observations or catalogs using the recommended search radius for each instrument. We found MAXI observations of the GC at the beginning and end of the day on the 12th that show no significant variability or flaring. The 3$\sigma$ upper limit on the daily light curve flux in the 4--10\,keV energy range is 10$^{-9}$ erg\,s$^{-1}$\,cm$^{-2}$, which corresponds to a luminosity of  $\sim8\times10^{36}$ ergs\,s$^{-1}$ using a GC distance of 8.12\,kpc. If this source produced a flare, it must have faded in less than a day.

\section{Results and Discussion}\label{sec:results_disc}

\subsection{Luminosity Class of XID 6592}

\begin{figure}[ht]
  \centering
  \includegraphics[trim=10 10 44 36, clip,scale=0.55]{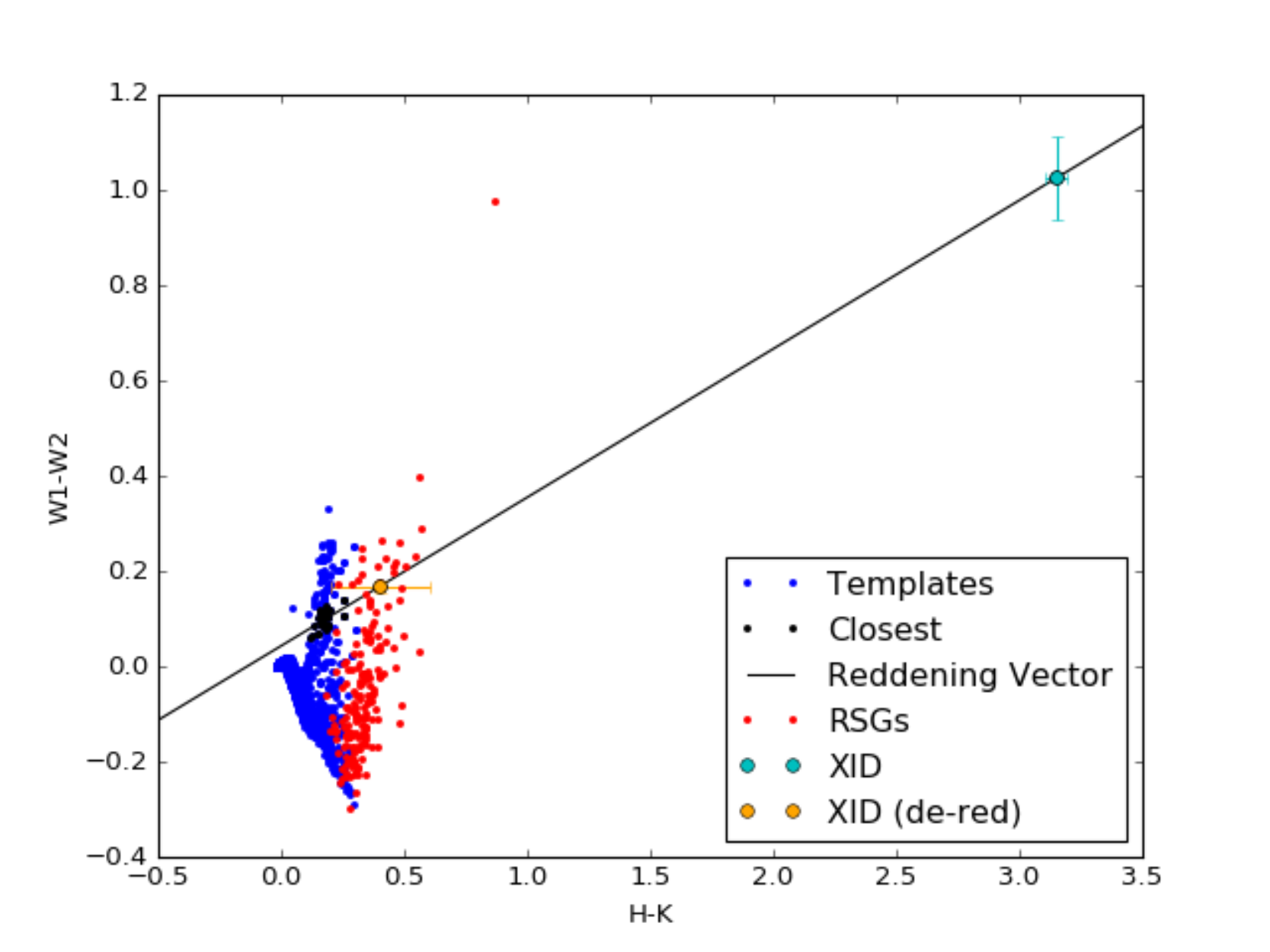}
  \caption{Color--color plot using $H, K_s$, and the first two WISE bands W1 and W2. The cyan data point is XID 6592's measured photometry, the black line is the reddening vector, the orange data point is XID 6592 dereddened by $\sim40$ mag of extinction in the $V$ band (which corresponds to $\sim4.5$ mag in $K_s$ band), the red points are photometry of known RSGs from \citet{Yang2011}, and the blue data points are photometry of synthetic stellar spectra from \citet{Coelho2014} obtained using a modified version of EzGal \citep{Mancone2012}.}\label{fig:xid_color} 
\end{figure}

In order to determine the luminosity class of XID 6592, we created a color-color plot (shown in Figure \ref{fig:xid_color}) using  IR photometry from WISE and ISPI and a modified version of EzGal \citep{Mancone2012} called EzMag (Jeram et al. 2020, in preparation) to obtain magnitudes of a variety of different stellar spectral templates from \citet{Coelho2014} with $T_{\mathrm{eff}}$ = 3000 -- 4250 K and log $g$ = --0.5--5.5 cgs, as well as known RSGs from \citet{Yang2011}.

\begin{figure}[ht]
  \includegraphics[trim=10 10 44 36, clip,scale=0.55]{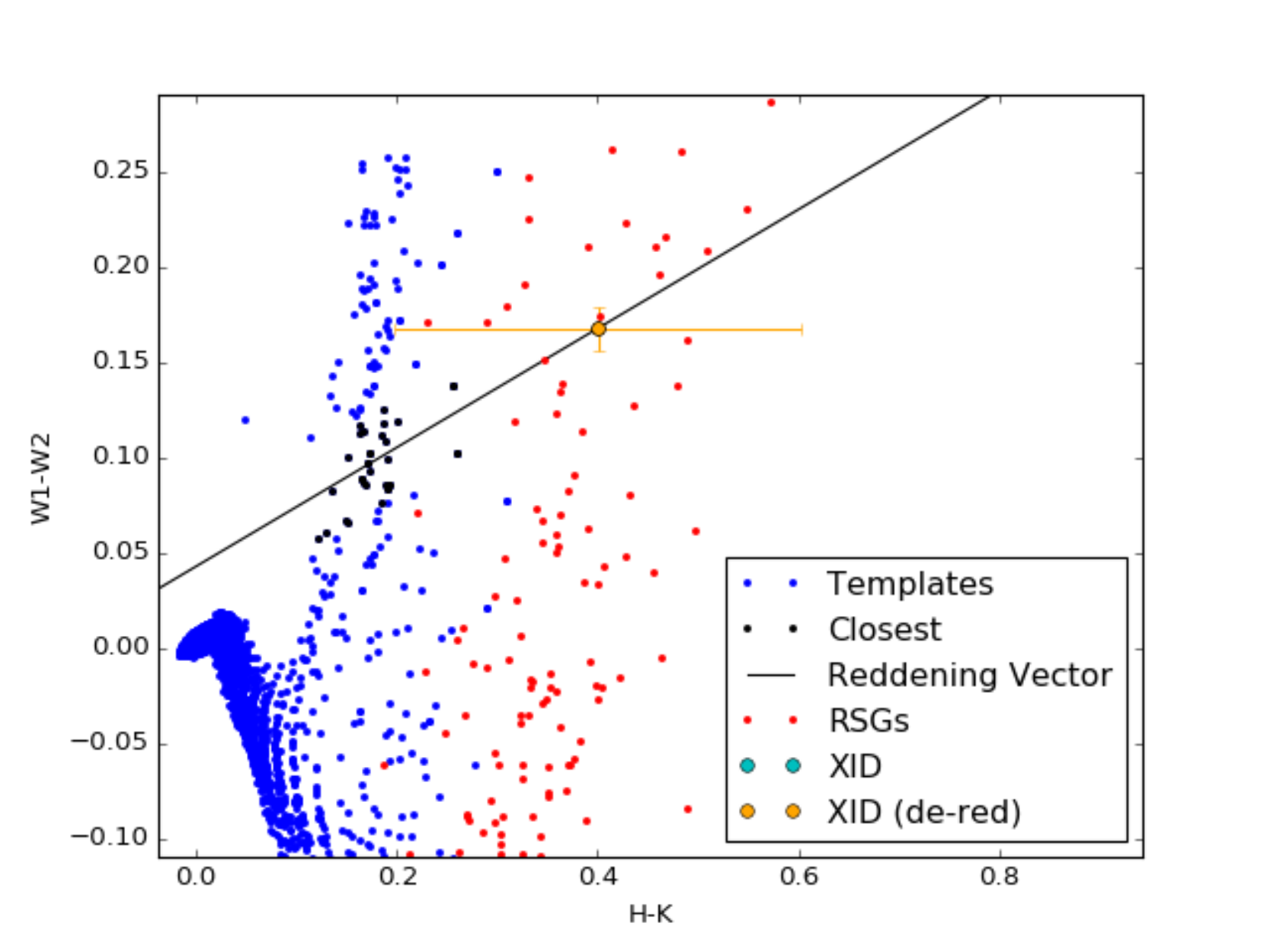}
  \caption{Zoom-in on Figure \ref{fig:xid_color}. 
  }\label{fig:xid_color_zoom}
\end{figure}

We expect XID 6592 to be located within the locus of known star colors, but it is far too red to match any template stars, due to the high extinction toward the GC. Instead of applying the extinction law of \citet{Nishiyama2006}, which preferentially relies on stars on the near side of the GC (Stelter et al. 2020, in press), we used the extinction law of \citet{Cardelli1989} to calculate the reddening vector. In order for XID 6592 to be located near known stars along this vector, the extinction in the $V$ band, $A_{\mathrm{V}}$, must be $39.5\pm3$ mag, corresponding to $A_{Ks} = 4.5\pm0.3$ mag. This extinction is higher than the typical GC value of $A_{Ks} \sim$3 mag \citep{GaoLi2013}. However, Stelter et al., (in prep) found that there is a broad range of extinction values with a bimodal distribution, peaking at $A_{Ks}$ of $\sim 4$ and $\sim$7.5 mag. Therefore, we conclude that XID 6592 may lie in a region of slightly higher-than-average extinction. We then calculated the absolute magnitude of XID 6592 to be $M_{Ks}=-9.4\pm0.3$ mag using this extinction value, a GC distance of $8.12\pm0.03$\,kpc, and an apparent magnitude of $m_{Ks}$ = $9.692\pm0.016$ mag (our CIRCE measurement).  As stated earlier, we assume that XID 6592 is at the GC because its location and extinction are consistent with the GC. There is also a much higher density of X-ray binaries in the GC than outside of it, making the probability of XID 6592 being in the foreground very low.

In Figure \ref{fig:xid_color_zoom}, the black data points are those that are closest to the reddening vector. These templates have colors of $H-K$ = 0.1--0.3, $T_{\mathrm{eff}}$ = 3000 - 4250 K, and log\,$g$ of -0.5 to 1 cgs and 4.5 to 5.5 cgs. However, the log $g$ of 4.5-5.5 corresponds to dwarf/main sequence stars, which are orders of magnitude fainter than XID 6592 in the $K$ band ($M_{Ks,\mathrm{XID}}$ = $-9.4\pm0.3$ mag), so we are left with $T_{\mathrm{eff}}$ = 3000--3400 K and log\,$g = -0.5$ to 1 cgs. While the spectral templates do not span the complete range of log $g$, this is supplemented by the photometry of known RSGs, and the bright absolute magnitude, color, and temperature are all consistent with an RSG.

\subsection{Infrared Fast Variability}\label{subsec:ir}

In this section, we discuss potential explanations for the significant fast IR variability observed in XID 6592. The variability we observe in the WISE 1 light curve is unusual for RSG stars, as they should not be capable of varying on short timescales at this amplitude. RSGs (later than spectral type M0) have very large radii ($\gtrapprox500$\,R$_{\odot}$; \citet{Cox2000}; K-type SGs have radii $\sim200-500$\,R$_{\odot}$), so any variation would take $\sim60$\,light-minutes to propagate across the photosphere. As a quick check, we calculated the percentage change in radius that would be required to cause the change in brightness that we observe. A change in magnitude of 0.5 corresponds to a change in luminosity of $\sim50\,\%$. If this is due to a change in radius, it would require a change in radius of 25\%. Cooler starspots also cannot account for the variation we see, because while they can cover a decent amount of the star's surface, they only cause brightness variations of $\sim1\%$, whereas we see brightness variations of $\sim50\%$. RSGs have observed photometric periodicities of $\sim100-1500$ days with amplitudes of $\sim0.5-6$ mag in the $K_s$ band \citep{Kiss2006}. If we calculate a rate of change (how many magnitudes these stars vary per day), we find that RSGs vary by $\sim0.01$ mag day$^{-1}$ compared with XID 6592, which varies at a rate of $\sim5$ mag day$^{-1}$; this rate is a factor of $\sim1000$ larger, which implies that the variability is not intrinsic to the RSG. In short, we do not know of any intrinsic physical mechanisms that would cause this large of a change in an RSG over such a short period of time, which implies that it must be produced by the compact object.

Because we know that this source is an X-ray binary, the IR variability could be caused by the compact object, which we assume is producing the X-ray emission from the system. One class of explanations is that the compact object is either ionizing the RSG’s wind or directly driving an outflow. The high velocity seen in the Br $\gamma$ line is more consistent with an outflow driven from an accretion disk around the binary companion than by the ionization of the wind from the RSG. In X-ray binaries, the observed IR variability is typically driven by X-ray-emitting accretion processes. In this scenario, some fraction of the X-rays produced by the accretion disk around the compact object will interact with the companion star and be reprocessed into infrared photons. However, this requires that the X-ray luminosity be larger than the IR variability luminosity - in other systems the X-ray luminosity is typically 10--1000$\times$ larger than the IR luminosity, while in XID 6592 this ratio is $\sim 0.001$. 

Alternately, this source could also be a neutron star symbiotic X-ray binary or a hard spectrum white dwarf symbiotic binary as discussed in \citet{DeWitt2013}, or the IR variability could be produced by a relativistic jet. We compared the XID 6592 IR flux with that of the black hole binary GX 339--4 \citep{Gandhi2011}, which contains a jet outflow. GX 339--4 changes by $\sim0.9$ mag over 6 hr, which is approximately the same variability timescale and (relative) amplitude as XID 6592. However, at $\sim10^{38}$ ergs\,s$^{-1}$, its X-ray luminosity is orders of magnitude larger than XID 6592. Furthermore, the ratio of X-ray to IR luminosities $L_{\mathrm{X-ray}}$/$L_{\mathrm{IR}}\sim$100 for GX 339-4, compared to  0.001 for XID 6592. In microquasars such as SS 433 \citep{Fabrika2004}, relativistic jets can cause large IR and quick variations and have an IR excess. In these systems, the IR luminosity can match and even exceed the X-ray luminosity. According to \citet{Russell2008}, the IR flux is related to the X-ray flux by F$_{\mathrm{IR}}\propto$ F$_{\mathrm{X-ray}}^{\alpha}$,where the power-law index $\alpha$ is $\sim -0.6-0.7$ for sources containing jets. However, $\alpha \sim -0.3$ for XID 6592, which is not consistent with a jet.

\subsection{X-Ray Fast Variability}\label{subsec:xray}

After analyzing archival Chandra and XMM observations, we find variability in the X-rays on timescales of days with amplitudes $5\times$ the quiescent luminosity in Chandra, as well as multiple points in time where the source went below the detection limit. We also see brightening events in XMM on timescales of tens to hundreds of days. The weighted average luminosity seen in the XMM data, $(1.81 \pm 0.03 )\times10^{33}$\,ergs\,s$^{-1}$, is not completely consistent with the luminosity seen by Chandra, $(1.27 \pm 0.07 )\times10^{33}$\,ergs\,s$^{-1}$. However, XMM has a broader energy range (0.2--12\,kev) compared to Chandra (0.2--8\,keV), and XID 6592 is more luminous at higher energies, which may explain this apparent discrepancy. 

This source was also observed by NuSTAR in \citet{Nustar2016} (source 63). They obtained $420 \pm 61$ net counts in the 3--40\,keV energy range with a hardness ratio of $0.02 \pm 24$ and a photon index of $\Gamma = 0.69 \pm 36$ assuming an $N_{\mathrm{H}}$ of $6\times10^{22}$\,cm$^{-2}$. The calculated fluxes and luminosities are shown in Table \ref{table:nustar}. The source is very hard in X-rays, meaning that there are more photons at higher energies than lower energies, but its flux at low energies is consistent with their Chandra 2--8\,keV flux of $13.3\times10^{-6}$\,photons\,s$^{-1}$\,cm$^{-2}$.

\begin{table}[]
\centering
\caption{NuSTAR results from \citet{Nustar2016}}
\label{table:nustar}
\begin{tabular}{ccc}
\toprule
Energy Range & Flux & Luminosity \\
(keV) & ($\times10^{-6}$ photons\,s$^{-1}$\,cm$^{-2}$) & ($\times10^{32}$ ergs\,s$^{-1}$) \\
\hline
3--10 & 11.5 (4.7) & 9.0 (3.7) \\
10--40 & 25.7 (4.2) & 72 (12) \\ 
\hline
\end{tabular}
\end{table}

The X-ray variability in the Chandra data comes from the compact object (which could be either a black hole or a neutron star). There are several possible mechanisms that could explain the X-ray variability. For instance, fluctuations in the accretion disk, potentially due to a variable mass accretion rate, could cause the X-ray variations. However, comparing the IR and X-ray luminosities, we find that the IR variability luminosity of $2\times10^{36}$\,ergs\,s$^{-1}$ is $\sim$1000$\times$ brighter than the maximum measured X-ray luminosity of $\sim$3$\times10^{33}$\, ergs\,s$^{-1}$. Even if this source reached the MAXI upper limit of $L_\mathrm{X-ray}\sim8\times10^{36}$\, ergs\,s$^{-1}$, this is still not enough to power the IR variability given reasonable X-ray reprocessing efficiencies. This implies that X-ray reprocessing is not responsible for the IR flux variations. 

\subsection{A Possible SFXT?}\label{subsec:sfxt}

Another possible explanation is that this system is a super (giant) fast X-ray transient (SFXT). SFXTs can be in quiescence for the majority of their lifetimes. They can have very low duty cycles (percentage of time spent as bright in X-rays) -- as low as 0.1\%, with typical quiescent luminosities of $10^{33}-10^{34}$\,ergs\,s$^{-1}$. XID 6592's quiescent behavior matches fairly well with the SFXT behavior, as it has a similar luminosity and stays in quiescence for long periods of time. SFXTs typically host neutron stars as the compact object. They also exhibit flares that have a large dynamic range: they can span $10^{2-6}\times$ the quiescent luminosity, sometimes reaching $10^{38}$\,ergs\,s$^{-1}$ (which would be needed to explain the IR variability in XID 6592). These flares are relatively short and typically last 10--10,000\,s \citep{Sidoli2017}.

One of the current theories of the physical mechanism behind SFXTs is the gating mechanism proposed by \citet{Bozzo2008} and \citet{Grebenev2007}. In this model, neutron stars/pulsars with very slow spin periods ($>1000$\,s) and very high, magnetar-like magnetic fields ($10^{14}$\,G) prevent accretion onto the neutron star via a magnetic barrier. Without material accreting onto the neutron star, the X-ray luminosity is low. However, an X-ray flare can occur if a clump of material from the wind of the supergiant is dense enough that it overcomes the barrier and manages to be accreted onto the neutron star -- though it is still unclear how or why this works. This is not the only explanation, as there are also cases where accretion disks exist around magnetars \citep{Zhang2010,Bernardini2013,Tong2015}.  One caveat of this theory is that the companion stars in SFXTs are typically O/B-type stars, whereas we clearly have a red M/K-type star in XID 6592. Another caveat is that the Chandra brightening events we see in XID 6592 reach only 6$\times$ the quiescent luminosity and last for much longer (0.5 days and $\sim50$ days). However, it is entirely possible that we may have missed an outburst from XID 6592 if/when it went into a much brighter but shorter outburst. If it reached $10^{38}$\,ergs\,s$^{-1}$ when the source changed by $\sim0.5$\,mag in the WISE 1 band, this would explain the very fast IR variability we see. However, as stated in Section \ref{subsec:xray_obs}, we do not see any significant flares in the X-rays at the beginning or end of this day. Therefore, if there was a flare, it must have faded on a timescale of hours, which is consistent with SFXT flares.

\subsection{Does XID 6592 really contain an RSG donor?}\label{subsec:rsg}

In order for the color of XID 6592 to match any stars in the $H-K_s$ and W1--W2 color plane (e.g., Figures \ref{fig:xid_color} and \ref{fig:xid_color_zoom}), this requires more than the currently accepted average extinction in the GC at the location of XID 6592. This is plausible because as Stelter et al., (2020, in preparation) found, the extinction in the GC is clumpy and varies across the GC, so it is plausible that XID 6592 could lie in a region of higher-than-average extinction. However, it is also possible that this source lies in a region of average extinction but with another angularly unresolved bright source very close by that has an even redder color (e.g., a red giant). While the GC region is very crowded, any unresolved source bright enough to change the observed colors would also have to participate in the large-amplitude variability. We are unaware of any physical scenario consistent with that, leaving us with the RSG scenario. 

We can consider the possibility that XID 6592 is a foreground red giant star. For a typical absolute magnitude of  $M_{Ks}$ = -1 mag \citep{Salaris2002} and an extinction of  $A_{Ks}$ = 4.5 mag (derived from the NIR and MIR colors above), we would arrive at a distance estimate of $\sim$100 -- 250 pc. However, interstellar extinctions do not reach such high levels over such short distances. To have such a high extinction, XID 6592 would have to be embedded in a (previously unknown) dense molecular cloud. Not only is such a cloud not known to exist at this location, but it would also be highly improbable to have a highly evolved star like a red giant embedded in such a molecular cloud core - such a situation is not known elsewhere in the Galaxy. Thus, while this scenario would significantly impact the IR and X-ray luminosities we discuss above in relation to the variability of XID 6592, we consider the possibility that XID 6592 is a foreground red giant star to be very unlikely.

Another possible explanation for the donor star is a carbon asymptotic giant branch (AGB) star, the most luminous of which reach $M_{Ks} >$ -8.0 mag \citep{Held2010}. At $M_{Ks}$ = -9.4 mag, XID 6592 is more luminous than this by 1.4 mag (i.e. a factor of 4), so for this source to be an AGB star, it would require some veiling by an accretion disk emission component. Oxygen AGB stars have similar colors to RSGs \citep{Wood1983}, and there are super-AGB stars that have similar brightnesses to RSGs. However, \citet{Doherty2017} state that it is generally difficult to distinguish the super-AGB stars from the RSGs. Therefore, we will simply include the super-AGB scenario as an option alongside the RSG scenario.

It is also possible that this source is some other type of object that exhibits fast IR variability, but XID 6592 has shown a Br $\gamma$ emission line in earlier observations, which is indicative of an accretion disk. It is also a variable hard X-ray source. Thus, the coincidental overlap of two such rare objects has a very low probability. Therefore, while it is possible that this source could be something other than an RSG in an X-ray binary, given our results, it seems unlikely.

Scutum X-1 is an X-ray binary system within our Galaxy that contains a late-type giant or supergiant and a neutron star with a 112\,s pulse period \citep{Kaplan2007}, and its properties are very similar to XID 6592 in terms of absolute magnitude, X-ray luminosity, and variability. The late-type companion is both very bright ($K_s$ = 6.55) and red ($J-K_s$ = 5.51), and the closest spectral type match is with late K to early M stars of luminosity classes I--III. The X-ray luminosity of Sct X-1 in the 0.5--10\,kev band is $1.4\times10^{33} d_{\mathrm{kpc}}^{2}$\,ergs\,s$^{-1}$ ($d\gtrsim$ 4\,kpc), the same order of magnitude as XID 6592. Furthermore, the overall X-ray flux has decreased by a factor of 4 over $\sim15$ yr, and over several months the pulsed amplitude varied by as much as a factor of 10 (XID 6592 shows X-ray variability of a factor of 6 across tens of days). Given the similarities between XID 6592 and Sct X-1, it is possible that XID 6592 may also host a neutron star compact object with a late-type supergiant companion.

RSG donors exist outside our Galaxy as well. \citet{Lau2019} searched for Spitzer/IRAC mid-IR counterparts of ultraluminious X-ray sources (ULXs), which are thought to be X-ray binaries with luminosities equal to or above the Eddington luminosity of a stellar-mass black hole. This luminosity is powered by accretion onto a compact object. \citet{Lau2019} found 12 counterparts with SG-like fluxes, where 4 sources were `red' ([3.6]--[4.5] $\sim$ 0.7) and 5 were `blue' ([3.6]--[4.5] $\sim$ 0) based on the IRAC colors. These correspond to sgB[e] and RSG companions, respectively. They compared infrared spectral energy distributions (SEDs) to SG template SEDs to further confirm the classification of spectral types. The remaining 3 companions were not previously detected in the mid-IR and are variable. After correcting for extinction, the [3.6]--[4.5] color of XID 6592 is consistent with the `bluer' objects, or the RSGs. XID 6592 could be a ULX that is heavily shrouded similar to SS 433, where the source has a luminosity of $\sim10^{38-39}$\,ergs\,s$^{-1}$ and X-ray reprocessing is still occurring, but we only see a small fraction of the total X-ray luminosity because the emission is not beamed toward us.

\section{Summary and Conclusions}\label{sec:summary}

The X-ray binary system XID 6592 is like no system we have seen before: it shows fast variability of $\sim$0.5 mag in only a few hours in the IR in the WISE 1 band. We see slower variability in VVV catalog, and none seen in our VVV light curve. We also see significant variability in the X-rays in both Chandra and XMM. We hypothesize that this system is an SFXT, which would make it the first SFXT to have an RSG for a companion star. However, more observations are needed to determine what part of the system is causing the large, fast IR variations, and in general the physical characteristics of the system. This can be accomplished through spectroscopy. We have observations from EMIR (Especrografo Multiobjeto Infra-Rojo; \citet{EMIR}) and FLAMINGOS-2 \citep{Eiken2012}, where we will search for spectral variability. We plan to determine which part of the spectrum is varying: the continuum, the Br$\gamma$ line (which would indicate variations in the accretion disk), or the CO bands (which would indicate that the companion star is varying).

\acknowledgments
ACKNOWLEDGEMENTS: 

The authors thank Ignacio Negueruela for his valuable input and discussion, and Sarik Jeram for his help with his EzMag code. 

A.M. acknowledges support from the Generalitat Valenciana through the grant BEST/2015/242 and from the Ministerio de Educación, Cultura y Deporte through the grant PRX15/00030.

This research has made use of the NASA/ IPAC Infrared Science Archive, which is operated by the Jet Propulsion Laboratory, California Institute of Technology, under contract with the National Aeronautics and Space Administration.

This publication makes use of data products from the Wide-field Infrared Survey Explorer, which is a joint project of the University of California, Los Angeles, and the Jet Propulsion Laboratory/California Institute of Technology, and NEOWISE, which is a project of the Jet Propulsion Laboratory/California Institute of Technology. WISE and NEOWISE are funded by the National Aeronautics and Space Administration.

Based on data products from observations made with ESO Telescopes at the La 
Silla or Paranal Observatories under ESO program ID 179.B-2002.

Based on observations made with the Gran Telescopio Canarias (GTC), installed in the Spanish Observatorio del Roque de los Muchachos of the Instituto de Astrofísica de Canarias, in the island of La Palma.

Development of CIRCE was supported by the University of Florida and the National Science Foundation (grant AST-0352664), in collaboration with IUCAA.

This publication makes use of data products from the Two Micron All Sky Survey, 
which is a joint project of the University of Massachusetts and the Infrared 
Processing and Analysis Center/California Institute of Technology, funded by the 
National Aeronautics and Space Administration and the National Science Foundation.

The scientific results reported in this article are based in part on data 
obtained from the Chandra Data Archive (ObsID 1561, 2291, 2293, 4683, 4684, 5950, 5951, 7037)

This research has made use of software provided by the Chandra X-ray Center (CXC) in the application package CIAO.

Based on observations obtained with XMM-Newton, an ESA science mission with 
instruments and contributions directly funded by ESA Member States and NASA. \\

\vspace{5mm}
\facilities{CTIO: 2MASS, WISE, Spitzer: IRAC, ESO: VISTA, GTC, CXO (ACIS), XMM}\\

\software{astropy \citep{2013A&A...558A..33A},  
          SExtractor \citep{1996A&AS..117..393B},
          SuperFATBOY  \citep{Warner2012,Warner2013},
          CIAO \citep{Fruscione2006}
          }

\end{document}